\documentclass{ws-procs10x7}
\usepackage{amssymb}
\usepackage[mathcal]{euscript}
\usepackage{epic}
\usepackage{eepic}
\usepackage{times}
\usepackage{epsfig,graphics}

\def\1{\one}
\def\2{\two}

\def\H3p{H_3^+}

\newcommand{\SU}{{\rm SU(2)}}

\newtheorem{rem}{Remark}

\newcommand{\sst}{\scriptscriptstyle}

\newcommand{\beq}{\begin{equation}}
\newcommand{\eeq}{\end{equation}}
\renewcommand{\theequation}{\thesection.\arabic{equation}} 

\newcommand{\id}{{\rm id}}

\newcommand{\pa}{\partial}
\newcommand{\ot}{\otimes}
\newcommand{\ra}{\rightarrow}

\newcommand{\ti}{\times}

\newcommand{\fr}[2]{{\textstyle \frac{#1}{#2} }}

\newcommand{\bra}{\langle}

\newcommand{\ket}{\rangle}

\newcommand{\al}{a}

\newcommand{\be}{\beta}
\newcommand{\ga}{\gamma}
\newcommand{\Ga}{\Gamma}
\newcommand{\de}{\delta}
\newcommand{\De}{\Delta}
\newcommand{\ep}{\epsilon}

\newcommand{\si}{\sigma}

\newcommand{\vf}{\varphi}

\newcommand{\bw}{\bar{w}}
\newcommand{\bx}{\bar{x}}
\newcommand{\bz}{\bar{z}}

\newcommand{\CA}{{\mathcal A}}
\newcommand{\CB}{{\mathcal B}}
\newcommand{\CC}{{\mathcal C}}

\newcommand{\CF}{{\mathcal F}}
\newcommand{\CG}{{\mathcal G}}
\newcommand{\CH}{{\mathcal H}}

\newcommand{\CK}{{\mathcal K}}

\newcommand{\CM}{{\mathcal M}}

\newcommand{\CO}{{\mathcal O}}

\newcommand{\CT}{{\mathcal T}}

\newcommand{\CV}{{\mathcal V}}

\newcommand{\SA}{{\mathsf A}}
\newcommand{\SB}{{\mathsf B}}
\newcommand{\SC}{{\mathsf C}}

\newcommand{\SO}{{\mathsf O}}

\renewcommand{\SS}{{\mathsf S}}

\renewcommand{\SU}{{\mathsf U}}

\renewcommand{\sf}{{\mathsf f}}

\newcommand{\sll}{{\mathsf l}}

\newcommand{\FB}{{\mathfrak B}}

\newcommand{\one}{{\mathfrak 1}}
\newcommand{\two}{{\mathfrak 2}}

\newcommand{\BR}{{\mathbb R}}
\newcommand{\BH}{{\mathbb H}}

\newcommand{\BI}{{\mathbb I}}
\newcommand{\BC}{{\mathbb C}}
\newcommand{\BP}{{\mathbb P}}
\newcommand{\BS}{{\mathbb S}}

\newcommand{\BZ}{{\mathbb Z}}

\newcommand{\rf}[1]{(\ref{#1})}

\newcommand{\aufz}
{\begin{list}{$\bullet$}{\topsep0cm \itemsep0cm \parsep0cm}}
\newcommand{\eaufz}{\end{list}}
\begin{document}
\title{From Liouville theory to the quantum geometry of Riemann surfaces}
\author{J. Teschner}
\address{Institut f\"ur theoretische Physik\\                 
Freie Universit\"at Berlin,\\                        
Arnimallee 14\\                                    
14195 Berlin\\ Germany}

\maketitle
\abstracts{The aim of this note is to propose 
an interpretation for the full (non-chiral) correlation
functions of the Liouville conformal field theory
within the context of the quantization of spaces
of Riemann surfaces. \\[2ex]
Contribution to the proceedings of the International Congress
of Mathematical Physics, Lisbon 2003 
}

\section{Introduction}

Liouville theory is a conformal field theory in two dimensions
which has a classical limit described by the (euclidean) action 
\begin{equation}
S[\phi]\;=\;\frac{1}{\pi}
\int d^2 z \;
\big(\pa_w\phi\pa_{\bw}\phi +\pi\mu e^{2b\phi}\big).
\end{equation}
Understanding the corresponding quantum theory is an important problem
in mathematical physics for at least two reasons:
\begin{itemize}
\item Quantum Liouville theory provides the simplest example for a 
two-dimensional conformal field theory with continuous spectrum \cite{CT,TL}. 
It can therefore be regarded as a paradigm for a whole new class
of two-dimensional conformal field theories which are neither 
rational nor quasi-rational.
\item The quantized Liouville theory is related
to quantized spaces of Riemann surfaces.
This interpretation should provide the 
basis for a deeper understanding
of two-dimensional quantum gravity \cite{Pol} as well as a future theory of
three-dimensional quantum gravity (see e.g. \cite{Kr,KV} and references 
therein).
\end{itemize}
In the following note we will be mainly concerned with the second of these
two points. The expectation that quantum Liouville theory 
is related to the quantum geometry of
Riemann surfaces goes back to Polyakov's work
\cite{Pol} and was formulated more precisely in \cite{V,T}. 
This interpretation was recently given a solid ground
\cite{TT1,TT2}
by establishing a direct relation between the conformal blocks 
of quantum Liouville theory \cite{TL,TL2} and the space of states 
obtained by quantizing the Teichm\"uller spaces of Riemann surfaces 
\cite{Fo,Ka1,CF,Ka2,Ka3,Ka4}.

Our aim in the present note will be to elaborate further on 
the geometrical interpretation
of quantum Liouville theory by proposing a (partly conjectural)
interpretation of the full (non-chiral) correlation functions of quantum
Liouville theory within quantum Teichm\"uller theory.

\section{The Liouville conformal field theory}

Quantum Liouville theory is a conformal field theory.
The space of states decomposes into irreducible representations
of the (left/right) Virasoro algebras as \cite{CT,TL}
\begin{equation}\label{spec}
\CH\;=\;\int_{\BS}d\al\;\CV_{\al,c}\ot\CV_{\al,c},\qquad\BS=\frac{Q}{2}+i\BR^+,
\end{equation}
where $Q=b+b^{-1}$ and $\CV_{\al,c}$, 
$\al\in\BS$ are the irreducible unitary 
representations of the Virasoro algebra
with central charge $c=1+6Q^2$ and 
highest weight $\De_{\al}=\al(Q-\al)$.
Being a conformal field theory, quantum Liouville theory is 
fully characterized by the set of s-point functions 
\begin{equation}\label{s-point}
\big\bra V_{\al_s}(z_s,\bz_s)\dots V_{\al_1}(z_1,\bz_1)\big\ket
\end{equation}
of the primary fields $V_{\al}(z,\bz)$, $\al\in\BC$
with conformal dimensions $\De_{\al}=\al(Q-\al)$. The 
M\"obius-invariance of the s-point functions allows us to assume
$z_s=\infty$, $z_{s-1}=1$ and $z_{s-2}=0$.

The primary fields $V_{\al}(z,\bz)$ 
of quantum Liouville theory were constructed in
\cite{TL,TL2}. With the help of the 
constructions in \cite{TL,TL2} it is possible to show that the 
s-point functions can be represented in a holomorphically 
factorized form
\begin{equation}\label{s-point2}\begin{aligned}
\big\bra & V_{\al_s}(z_s,\bz_s)\dots V_{\al_1}(z_1,\bz_1)\big\ket\;=\;
& \int_{\BS^{\kappa}_{}}dS\,m(S)\;\CF_{S,A}(Z)\,
\CF_{S,A}
(\bar{Z})\, .
\end{aligned}
\end{equation}
In order to write \rf{s-point2} compactly we have introduced  
the tuples of variables $A=(\al_1,\dots,\al_s)$, 
$S=(\be_1,\dots,\be_{\kappa})$,
$Z=(z_1,\dots,z_{\kappa})$
and $\bar{Z}=(\bz_1,\dots,\bz_{\kappa})$, where $\kappa=s-3$. 
The tuple $S$ is integrated over
$\BS^{\kappa}$, where $\BS=\frac{Q}{2}+i\BR^+$,  
and the measure $m(S)$ is given as
\begin{equation}\label{measure}
m(S)\;=\;
\prod_{i=1}^{\kappa}\,4\sin\pi b(2\be_i-Q)\sin\pi b^{-1}(Q-2\be_i).
\end{equation}
The key objects in \rf{s-point2} are the {\it conformal blocks}
$\CF_{S,A}(Z)$. In the remainder of this section, adapted 
from \cite{TT1}, we will briefly describe the definition 
and some relevant properties of the
conformal blocks.

\subsection{The conformal Ward identities}

It is well-known that the conformal blocks are strongly 
constrained by the conformal Ward-identities which express
the conservation of energy-momentum on the punctured Riemann-sphere
$\Sigma\equiv{\mathbb P}^1\setminus\{z_1,\dots,z_s\}$. 
In order to exhibit the mathematical
content of the conformal Ward identities 
let us consider functionals
\[
\CF_{\rm\sst A}^{\Sigma} :
\CV_{\al_s}\ot\dots\ot\CV_{\al_1}\rightarrow \BC\] 
that satisfy the 
following invariance condition. Let $v(z)$ be a 
meromorphic vector field that is
holomorphic on ${\mathbb P}^1\setminus\{z_1,\dots,z_s\}$.
Write the Laurent-expansion of $v(z)$ 
in an annular neighborhood of $z_k$ 
in the form $v(z)=\sum_{n\in\BZ}v_n^{(k)}(z-z_k)^{n+1}$, and define
an operator $T[v]$ on $\CV_{\al_s}\ot\dots\ot\CV_{\al_1}$
by 
\renewcommand{\id}{{\rm id}}
\[
T[v]\;=\;\sum_{k=1}^s\sum_{n\in\BZ}v_n^{(k)}L_n^{(k)},\qquad
L_n^{(k)}=\id\ot\dots\ot
\underset{\scriptstyle\rm (k-th)}{L_n}\ot\dots\ot\id.
\] 
The conformal Ward identities can then be formulated as 
the condition that 
\begin{equation}\label{Wardid}
\CF_{\rm\sst A}^{\Sigma}\big(\,T[v] w\,\big)\;=\;0 
\end{equation}
holds for all $ w\in  \CV_{\al_s}\ot\dots\ot\CV_{\al_1}$ and all
meromorphic vector fields $v$ that are holomorphic 
on $\Sigma$. 

By choosing vector fields $v$ 
that are singular at a single point only one 
gets rules for moving Virasoro generators from one puncture 
to the other ones. In this way one may convince oneself that the 
functional $\CF_{\rm\sst A}^{\Sigma}$
is uniquely determined by the value 
$\CF_{S,A}(Z)\equiv
\CF_{\rm\sst A}^{\Sigma}(v_{\rm\sst A})\in\BC$ that it takes 
on the product of highest weight states
$v_{\rm\sst A}\equiv v_{\al_s}\ot\dots\ot v_{\al_1}$.


It is well-known that the space of solutions to the 
condition (\ref{Wardid}) is one-dimensional for 
the case of the three-punctured sphere $s=3$. Invariance under global
conformal transformations allows one to assume that
$\Sigma_3=\BP^1\setminus\{0,1,\infty\}$. 
We will adopt the normalization from \cite{TL} and denote
$\CC^{\Sigma_3}_{\rm\sst A}$ the unique conformal block 
that satisfies 
$\CC^{\Sigma_3}_{\rm\sst A}(v_{\al_3}\ot v_{\al_2}\ot v_{\al_1})=
{\rm N}(\al_3,\al_2,\al_1)$. The function ${\rm N}(\al_3,\al_2,\al_1)$
is defined in \cite{TL} but will not be needed explicitly
in the following.

Let us furthermore note that the case of $s=2$
corresponds to the invariant bilinear form 
\renewcommand{\bra}{\langle}
\renewcommand{\ket}{\rangle}
$\bra .\, , .\ket_{\al}:
\CV_{\al}\ot\CV_{\al}\rightarrow \BC$ which is defined 
such that $\bra L_{-n} w, v\ket_{\al}=
\bra  w,L_{n} v\ket_{\al}$. 

\subsection{Sewing of conformal blocks}\label{sewing}

For $s>3$ one
may generate large classes of solutions 
of the conformal Ward identities by the 
following ``sewing'' construction. Let 
$\Sigma_i$, $i=1,2$ be Riemann surfaces with $m_i+1$ punctures, 
and let $\CG_{\rm\sst A_2}^{\Sigma_2}$ and $\CH_{\rm\sst A_1}^{\Sigma_1}$
be conformal 
blocks associated to $\Sigma_i$, $i=1,2$ and representations
labeled by $A_2=(\al_{m_2},\dots,\al_1,\al)$ 
and $A_1=(\al,\al'_{m_1},\dots,\al'_{1})$ respectively.
Let $P_i$, $i=1,2$ be the distinguished punctures on $\Sigma_i$
that are associated to the representation $\CV_{\al}$. Around 
$P_i$ choose local coordinates $z_i$ such that $z_i=0$ parameterizes
the points $P_i$ themselves. Let $A_i$ be the annuli 
$|q|/R<|z_i|<R$ for  $R\in\BR^+$, $q\in\BC$, $|q|<R^2$, 
and let $D_i$ be the disks $|z_i|\leq |q|/R$.
We assume that $R$ is small enough such
that the annuli $A_i$ contain no other punctures.  
The surface $\Sigma_2\infty\Sigma_1$ 
that is obtained by ``sewing'' $\Sigma_2$ and $\Sigma_1$ will
be 
\[
\Sigma_2\,\infty\,\Sigma_1\;=\;\big(
(\Sigma_2\setminus D_2)\cup
(\Sigma_1\setminus D_1)\big)\,/\,{\sim},
\]
where $\sim$ denotes the identification of annuli $A_2$ and $A_1$
via $z_1z_2=q$. The conformal block 
$\CF_{\al,{\rm \sst A_{21}}}^{\Sigma_2\infty\Sigma_1}$ assigned to
$\Sigma_2\infty\Sigma_1$ , 
${\rm A}_{21}=(\al_{m_2},\dots,\al_1,\al'_{m_1},\dots,\al'_{1})$
and $\al\in\BS$ will then be
\begin{equation}\begin{aligned}
{}& \CF_{\al,{\rm \sst A_{21}}}^{\Sigma_2\infty\Sigma_1} 
\big(v_{m_2}\ot\dots\ot v_{1}\ot w_{m_1}\ot\dots \ot w_{1}\big)
\;=\;\\
 &=\sum_{\imath,\jmath\in\BI}\;
\CG_{\rm \sst A_2}^{\Sigma_2} 
\big( v_{m_2}^{}\ot\dots\ot v_{1}^{}\ot v_{a,\imath}^{}\big)
\,\bra  v^{\vee}_{a,\imath},q^{L_0} v_{a,\jmath}^{}
\ket_{\al}^{\phantom{\dagger}}\,
\CH_{\rm \sst A_1}^{\Sigma_1} 
\big( v^{\vee}_{a,\jmath}\ot w_{m_1}^{}\ot\dots\ot w_{1}^{}\big).
\end{aligned}
\end{equation}
The sets $\{ v_{\al,\imath};\imath\in\BI\}$ and
$\{ v_{\al,\imath}^{\vee};\imath\in\BI\}$ 
are supposed to represent mutually
dual bases for $\CV_{\al}$ in the sense that $
\bra \, v_{\al,\imath}^{\phantom{\vee}}, v_{\al,\jmath}^{\vee}\,
\ket_{\al}^{\phantom{\vee}}=\de_{\imath\jmath}$. In a similar way one 
may construct the conformal blocks associated to a surface that
was obtained by sewing two punctures on a single Riemann surface.

\subsection{Feynman rules for the construction of conformal blocks}

The sewing construction allows one to construct large classes of 
solutions to the conformal Ward identities from simple pieces. 
The resulting construction resembles
the construction of field theoretical amplitudes by 
the application of a set of Feyman rules.
Let us summarize the basic ingredients and their geometric 
counterparts. \\[1ex]
\newcommand{\gs}{$\frac{\quad}{}\;\;$}
{\sc Propagator} \gs Invariant bilinear form:
\[
\bra v_2\,,\,e^{-tL_0}v_1\ket_{\CV_{\al}}^{}\quad \sim\quad
\lower.8cm\hbox{\epsfig{figure=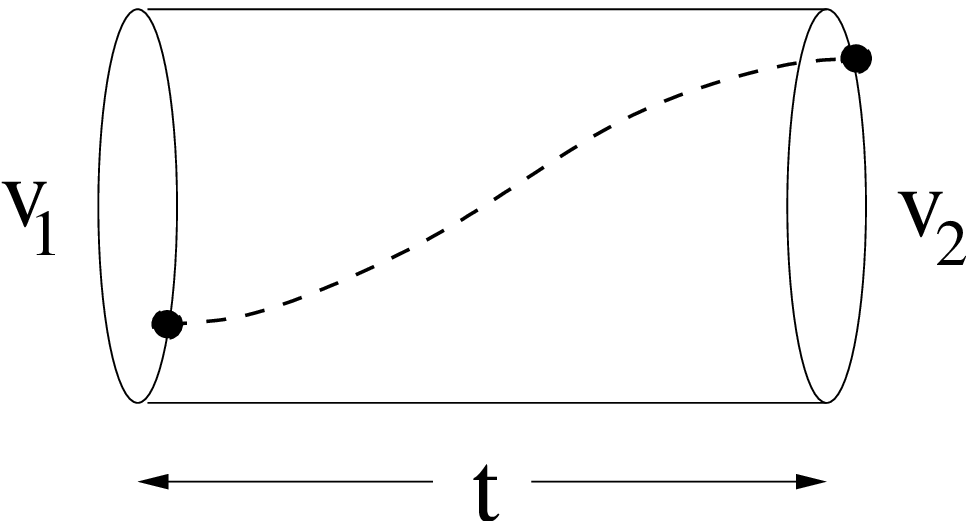,height=1.6cm}}
\]
{\sc Vertex} \gs Invariant trilinear form:
\[
\CC^{\Sigma_3}_{\rm\sst A}(v_3,v_2,v_1)\quad \sim\quad
\lower1.5cm\hbox{\epsfig{figure=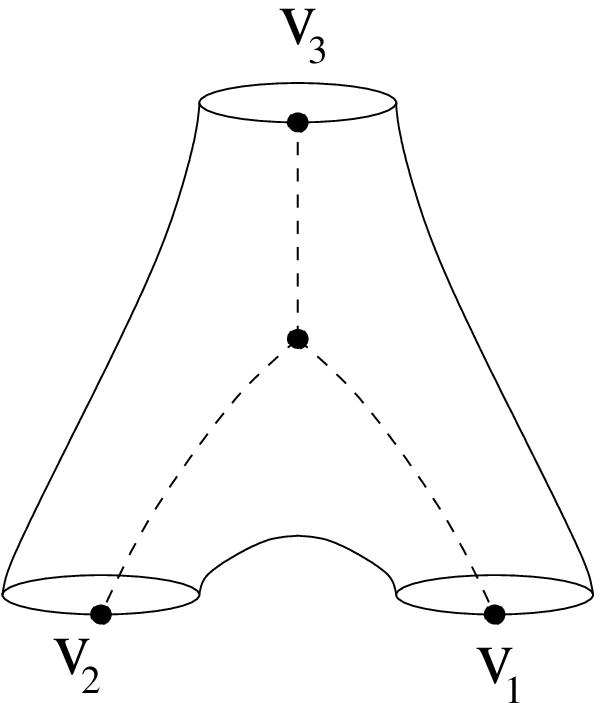,height=3cm}}
\]
{\sc Gluing} \gs Completeness: 
\[\begin{aligned}{}&
\lower0.7cm\hbox{\epsfig{figure=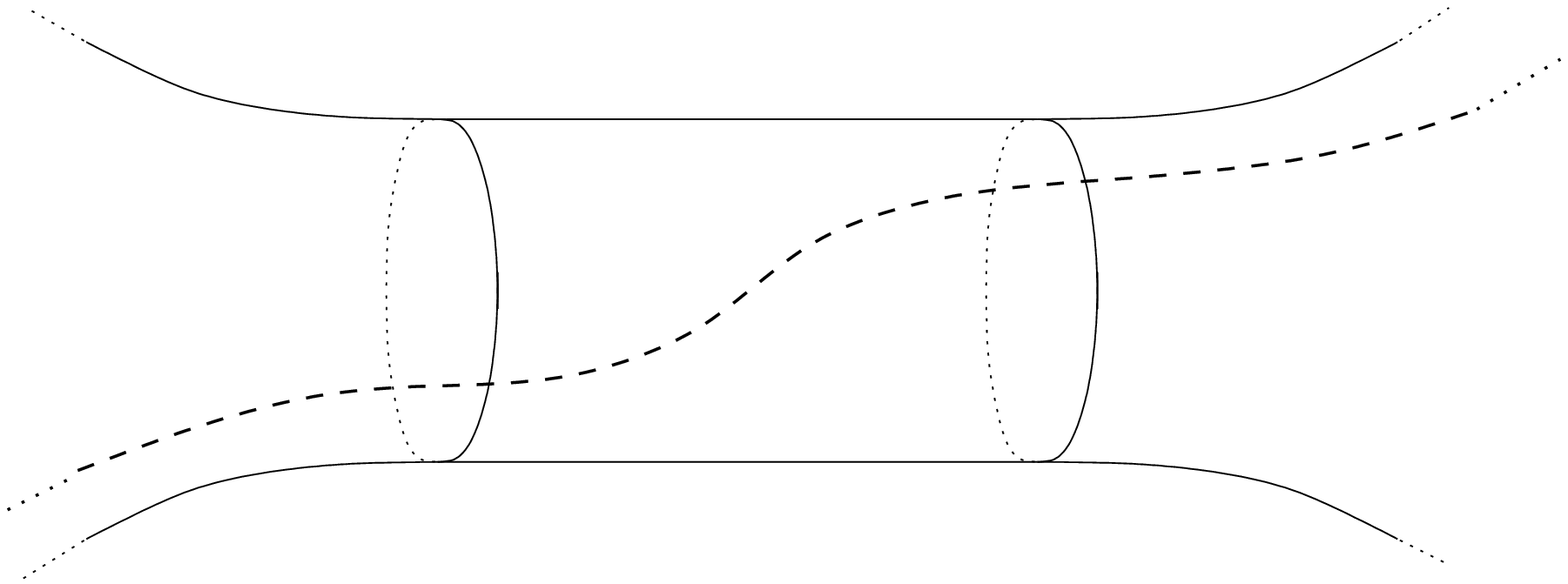,height=1.4cm}} \;=
\;\sum_{\imath,\jmath\in\BI}
\left(\,\;\lower0.7cm\hbox{\epsfig{figure=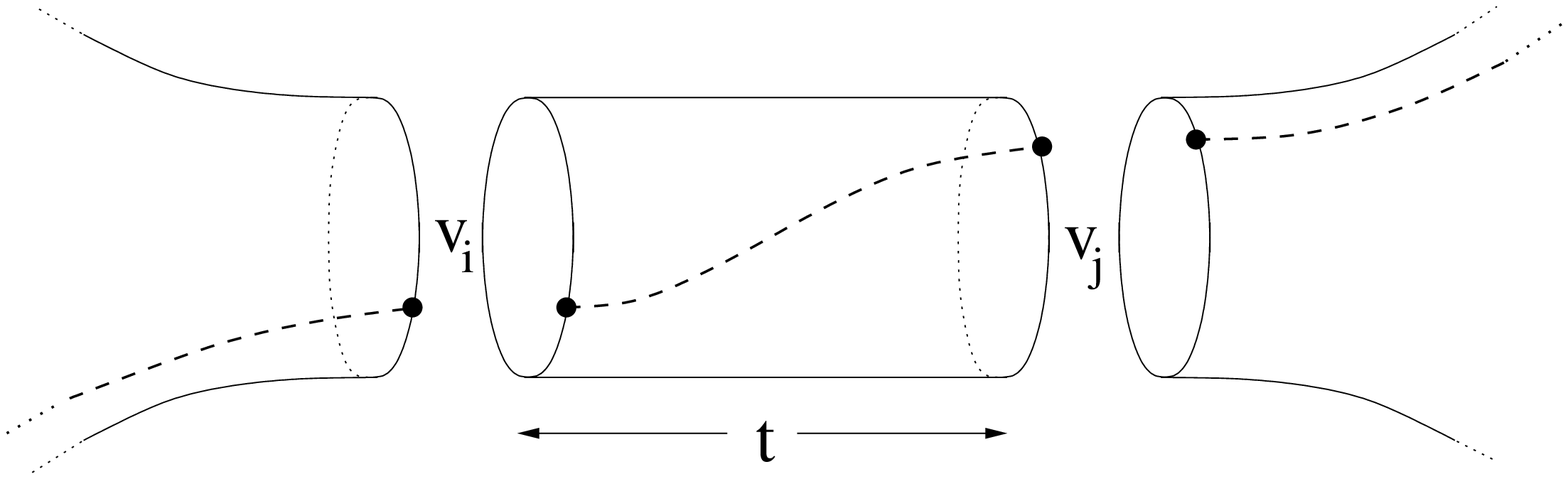,height=1.4cm}}\;\,\right)
\end{aligned}
\]
The variable $t$ is related to $q$ via $q=e^{-t}$.
We have  introduced the  dashed lines in order to take care of the 
fact that the rotation of a boundary circle by $2\pi$ (Dehn twist)
is not represented 
trivially. It acts by multiplication with $e^{2\pi i \De_{\al}}$. 
This describes a part of the action of the mapping class group
on the spaces of conformal blocks, which will 
be further discussed below. 
The Riemann surfaces that
are obtained by gluing cylinders and three-holed spheres as drawn will 
therefore carry a trivalent graph which we will call 
a Moore-Seiberg graph. 

The gluing construction furnishes  spaces of conformal blocks 
$\CH^{\rm L}(\Sigma,\Gamma)$ 
associated to 
a Riemann surface $\Sigma$ together with
a Moore-Seiberg graph $\Gamma$. A basis for this space is obtained
by coloring the ``internal'' edges of the Moore-Seiberg graph $\Gamma$
with elements of $\BS$, for example
\[
\CH^{\rm L}\left(
\lower.6cm\hbox{\epsfig{figure=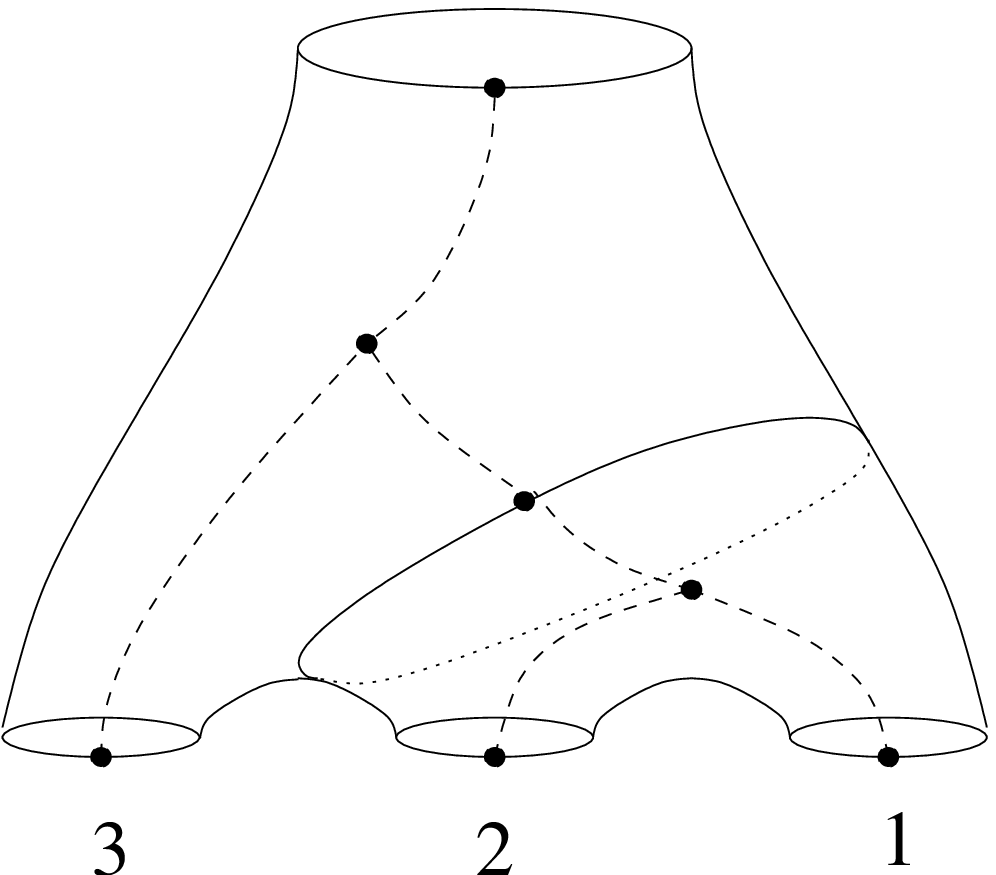,height=1.2cm}}
\right)\;= \;
{\rm Span}\left\{\;\;
\lower1.2cm\hbox{\epsfig{figure=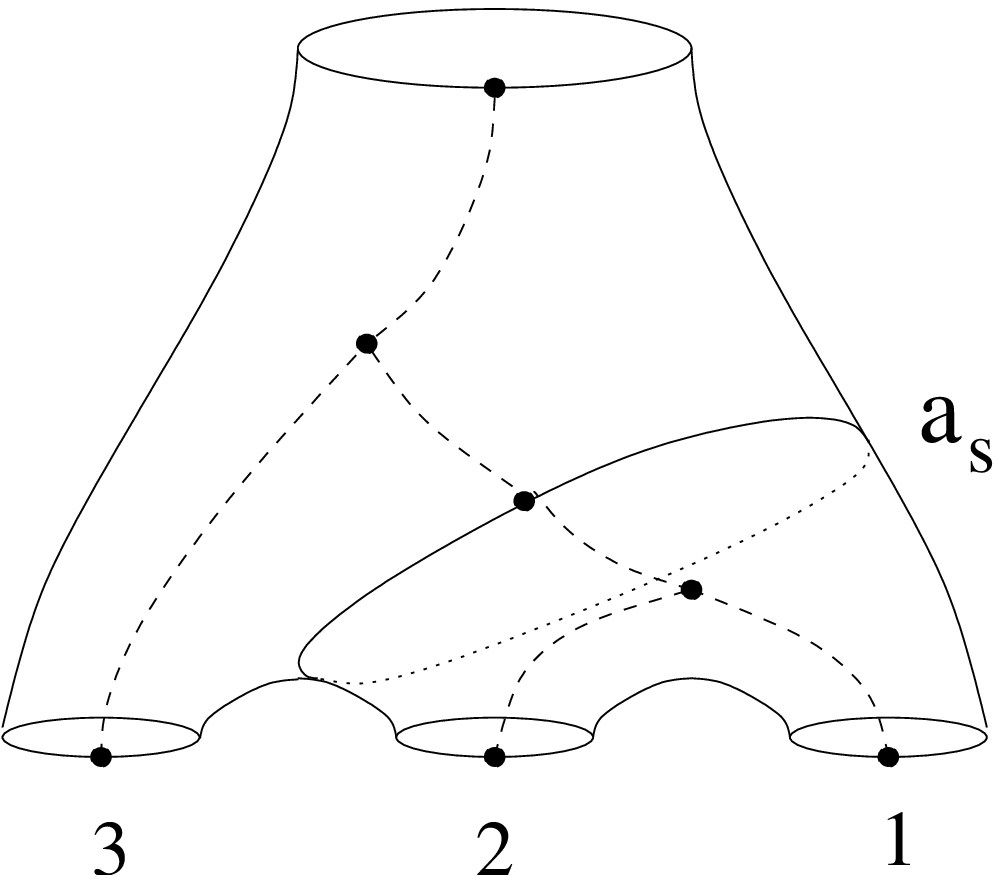,height=2.4cm}}
\;\; ;\; \al_s\in\BS\;\right\}.
\]
In order to show that the spaces of conformal blocks associated to 
each two Moore-Seiberg graphs $\Gamma_1$ and $\Gamma_2$ are 
isomorphic,
$\CH^{\rm L}(\Sigma,\Gamma_1)\simeq
\CH^{\rm L}(\Sigma,\Gamma_2)\simeq
\CH^{\rm L}(\Sigma),$
one needs to find operators 
$
\SU_{\Gamma_2\Gamma_1}^{}:
\CH^{\rm L}(\Sigma,\Gamma_1)\;\ra\;\CH^{\rm L}(\Sigma,\Gamma_2).$
We will describe the construction of such operators below.

\subsection{The Moore-Seiberg groupoid}

The transitions between different Moore-Seiberg graphs 
on a surface $\Sigma$ of genus zero with $s$ punctures generate
a groupoid ${\rm MS}^0_s$ that will be called the
Moore-Seiberg groupoid. This groupoid can be characterized
by generators and relations \cite{MS1,BK1}.
The set of generators for the groupoids ${\rm MS}^0_s$ 
associated to subsurfaces of genus zero 
is pictorially represented 
in Figure \ref{MS1a} below. 
\begin{figure}[h]
\centerline{\epsfxsize6cm\epsfbox{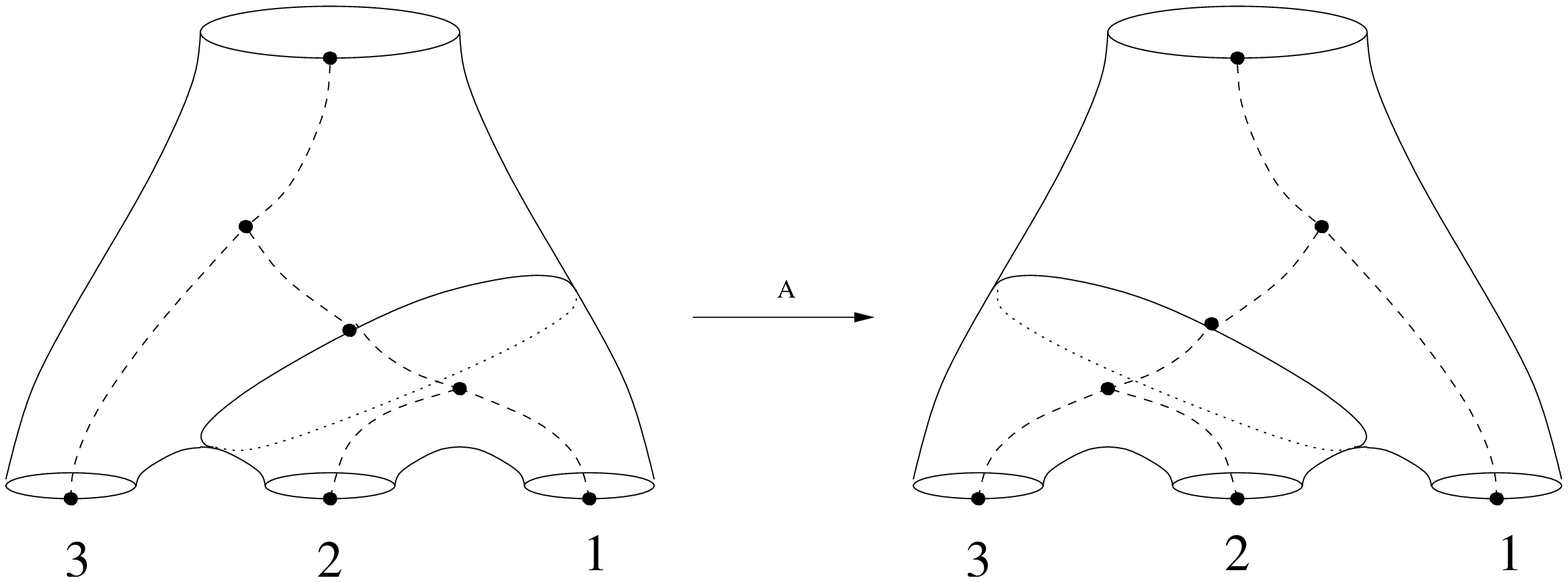}\hspace{1cm}
\epsfxsize6cm\epsfbox{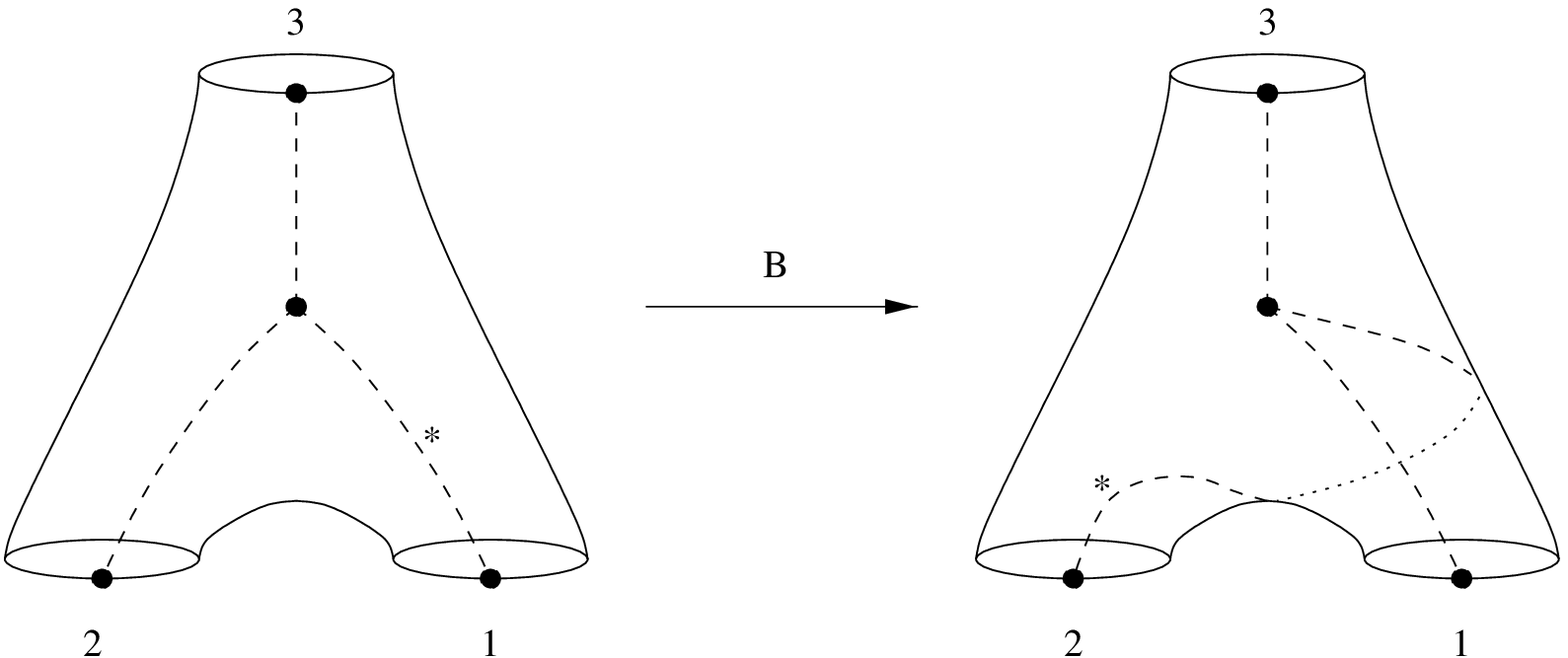}} 
\caption{The A- and B-moves}\label{MS1a}\end{figure}

One will get a relation in the Moore-Seiberg groupoid for 
every sequence of elementary transformations that leads back to the
same graph. Any such sequence will be identified with the 
trivial (identity) transformation.
However, all relations of the Moore-Seiberg groupoid can be
shown to follow from a finite set of basic relations \cite{MS1,BK1}.
In order to derive the basic relations in ${\rm MS}^0_s$
it suffices to consider the cases $s=4,5$.

\subsection{Representation of the Moore-Seiberg 
groupoid on $\CH^{\rm L}(\Sigma)$} 

In order to characterize a 
representation of the Moore-Seiberg groupoids ${\rm MS}^0_s$ it suffices 
to specify the operators 
$\SU_{\Gamma_2\Gamma_1}^{}$ for the cases where $\Gamma_2$ and $\Gamma_1$
differ by an A- or B-move. In the case of the Liouville 
conformal blocks in genus zero this was done in \cite{TL,TL2}.\\[1ex]
{\sc A-Move}: In order to describe the representation of the
A-move let $\Sigma$ be the 
four-punctured sphere, with parameters ${\rm A}=(\al_4,\dots,\al_1)$
associated to the four punctures respectively. The conformal blocks
corresponding to the two sewing patterns indicated on the
left half of Figure \ref{MS1a} will be denoted
$\CF_{\rm\sst A,\al_s}^{\Sigma}$ and $\CG_{\rm\sst A,\al_t}^{\Sigma}$
respectively, where $\CF_{\rm\sst A,\al_s}^{\Sigma}$ corresponds to
the leftmost part of Figure \ref{MS1a}.
The A-move is then represented as an integral
transformation of the following form:
\begin{equation}
\CF_{\rm A,\al_s}^{\Sigma}\;=\;\int_{\BS}d\mu(\al_t)\;
F_{\al_s\al_t}^{\rm\sst Liou}\big[\,{}_{\al_4}^{\al_3}\;{}_{\al_1}^{\al_2}\,\big]
\;\CG_{\rm A,\al_t}^{\Sigma}\; .
\end{equation}
The kernel 
$F_{\al_s\al_t}^{\rm\sst Liou}
\big[\,{}_{\al_4}^{\al_3}\;{}_{\al_1}^{\al_2}\,\big]$ 
is given by the following expression:
\begin{equation}\label{fuscoeff1}\begin{aligned}
F_{\al_s\al_t}^{\rm\sst Liou}
\big[\,{}_{\al_4}^{\al_3}\;{}_{\al_1}^{\al_2}\,\big]\;=\;
\frac{s_b(u_1)}{s_b(u_2)}\frac{s_b(w_1)}{s_b(w_2)}
\int\limits_{\BR}dt\;\,\prod_{i=1}^4\;\frac{s_b(t-r_i)}{s_b(t-s_i)},
\end{aligned}\end{equation}
where the special function $s_b(x)$ can be defined
by means of the following integral representation 
\begin{equation}
\log s_b(x)\;=\;\frac{1}{i}\int_{0}^{\infty}\frac{dt}{t}\left(
\frac{\sin 2xt}{2\sinh bt \sinh b^{-1}t}-\frac{x}{t}\right).
\end{equation}
In order to define the coefficients $r_i$, $s_i$, $u_i$ and $w_i$
let us introduce $c_b=i\frac{Q}{2}$ and 
write $\al_\flat$, $\flat\in\{1,2,3,4,s,t\}$, as 
$\al_\flat=\fr{Q}{2}+ip_{\flat}$.  
\begin{equation}  
\begin{aligned} r_\1=& p_2-p_1,\\
         r_\2=& p_2+p_1, \\
         r_3=& -p_4-p_3,\\
        r_4=& -p_4+p_3,
\end{aligned}\qquad\quad
\begin{aligned} 
       s_\1=& c_b-p_4+p_2-p_t,\\
         s_\2=& c_b-p_4+p_2+p_t,\\
         s_3=& c_b+p_s,\\
        s_4=& c_b-p_s , 
\end{aligned}
\qquad\quad \begin{aligned} u_\1=& p_s+p_2-p_1,\\
         u_\2=& p_s+p_3+p_4,\\
         w_1=& p_t+p_1+p_4,\\
        w_2=& p_t+p_2-p_3 , 
\end{aligned} \end{equation}
Setting $\al_t=\fr{Q}{2}+ip_t$ one 
may finally write the measure $d\mu(\al_t)$ in the form 
$d\mu(\al_t)=dp_t m(p_t)$, where 
$m(p_t)=4\sinh2\pi bp_t\sinh2\pi b^{-1}p_t$. 
\\[1ex]
\noindent
{\sc B-Move}: The B-move is realized simply by the multiplication 
with the phase factor
\begin{equation}
B^{\rm\sst Liou}(\al_3,\al_2,\al_1)\;=\;e^{\pi i(\De_{\al_3}-\De_{\al_2}-
\De_{\al_1})},
\end{equation}
where $\De_{\al_k}$, $k=1,2,3$ are the conformal dimensions
$\De_{\al}=\al(Q-\al)$.

\section{Relations between classical Liouville theory 
         and Teichm\"uller theory}

\subsection{The semi-classical limit}
In order to present first hints towards the geometrical 
interpretation of quantum Liouville theory let us consider 
a semi-classical limit of the Liouville correlation 
functions, following \cite{ZZ,TZ1}. We will study the limit 
when $b\ra 0$ with $\eta_i=b\al_i$, $i=1,\dots, n$ fixed. 
Noting that the rescaling 
\begin{equation}
\vf\;=\;2b\phi
\end{equation}
relates $\phi$ to the classical Liouville field $\vf$ and $b$ to 
$\sqrt{\hbar}$ one is lead to the expectation that the 
semi-classical limit of the correlation
functions should be of the form
\begin{equation}\label{classlim}
\big\bra V_{\al_s}(z_s,\bz_s)\dots V_{\al_1}(z_1,\bz_1)\big\ket
\;\underset{b\ra 0}{\ra}\;
\exp\big(-b^{-2}S^{\rm cl}[\vf]\,\big),
\end{equation}
where $\vf=\vf(z,\bz|E,Z)$, 
$E=(\eta_1,\dots,\eta_s)$, 
$Z=(z_1,\dots,z_s)$
is the unique solution
to the euclidean Liouville equation 
\begin{equation}\label{cleom}
\pa\bar{\pa}\vf\;=\;2\pi \mu_{\rm cl}^{} e^{\vf}, \qquad \mu_{\rm cl}^{}\equiv
\mu b^2,
\end{equation}
with the boundary conditions 
\begin{equation}\label{bcs}
\begin{aligned}
\vf (z,\bz)& =-2(1-\eta_s)\log|z|^2+\CO(1)\\
\vf (z,\bz)& =-2\eta_i\log|z-z_i|^2+\CO(1)
\end{aligned}\quad
\begin{aligned}
{\rm at}&\;\;|z|\ra z_s=\infty,\\
{\rm at}&\;\;|z|\ra z_i, \;i=1,\dots,s-1.
\end{aligned}
\end{equation}
The divergence of $\vf$ near the 
singular points $z_1,\dots,z_{s-1},\infty$ requires the 
inclusion of suitable regularization terms in the definition of 
the classical action $S^{\rm cl}$:
$S^{\rm cl}\big[\vf]=\lim_{\ep\ra 0}S^{\rm cl}_{\ep}\big[\vf]$, where
\begin{equation}
\begin{aligned}
S^{\rm cl}_{\ep}\big[\vf]\;=\; & \frac{1}{4\pi}
\int_{X^{\ep}}d^2z \,\Big(\,|\pa_z\vf|^2+\mu_{\rm cl} e^{\vf}\,\Big)\\
& \qquad -\sum_{i=1}^{s-1}
\bigg(\frac{\eta_i}{2\pi\ep}\int_{\pa D_i}dx \;\vf+2\eta_i^2\log\ep\bigg)\\
& \qquad +(1-\eta_s)
\bigg(\frac{\ep}{2\pi}\int_{\pa D_s} dx \;\vf-2\log\ep\bigg),
\end{aligned}
\end{equation}
where $D_i=\{z\in\BC;|z-z_i|<\ep\}$, 
$D_s=\{z\in\BC;|z|>1/\ep\}$, and 
$X^{\ep}=D_s\setminus\bigcup_{i=1}^{s-1} D_i$.

\begin{remark}
It is not rigorously proven yet that the Liouville correlation
functions constructed in the previous section have 
a semi-classical asymptotics given by \rf{classlim}. 
So far it was directly verified only in the case of the
three-point function \cite{ZZ}. Evidence for the validity
of \rf{classlim} for $s>3$ will follow from our
discussion in \S5.
\end{remark}

\subsection{Energy-momentum tensor and accessory parameters}

An important observable is the energy-momentum tensor $T(z)$.
In the classical Liouville theory it may be defined as
\begin{equation}\label{Tdef}
T_{\vf}(z)\;=\;-\fr{1}{2}(\vf_z)^2+\vf_{zz}.
\end{equation}
It is a classical result that the evaluation of $T_{\vf}(z)$ on the 
(unique) solution of equations \rf{cleom} and \rf{bcs} yields an expression 
of the following form:
\begin{equation}\label{Texp}
T_{\vf}(z)\;=\;\sum_{i=1}^{s-1}\left(\frac{\de_i}{(z-z_i)^2}+\frac{C_i}{z-z_i}
\right),
\end{equation}
where $\de_i=\eta_i(1-\eta_i)$. The asymptotic behavior of 
$T_{\vf}(z)$ near $z=\infty$ may be represented as
\begin{equation}\label{Tas}
T_{\vf}(z)\;=\;\frac{\de_s}{z^2}+\frac{C_s}{z^3}+O(z^{-4}).
\end{equation}
The so-called accessory parameters $C_i$,
$i=1,\dots, s$
are nontrivial functions on the moduli space
\begin{equation}
\CM_s^0\;=\;\big\{ (z_1,\dots,z_{s-3});z_i\neq 0,1\;\,{\rm and}\;\,
z_i\neq z_j\;\,{\rm for}\;\,i\neq j\,\big\} 
\end{equation}
of Riemann surfaces with genus 0 and $s$ punctures, 
which are restricted by the relations
\begin{equation}
\sum_{i=1}^{s-1}C_i=0,\qquad 
\sum_{i=1}^{s-1}(z_iC_i+h_i)=h_s,\qquad
\sum_{i=1}^{s-1}(z_i^2C_i+2h_iz_i)=C_s.
\end{equation}
It is instructive to compare \rf{Texp} to the 
conformal Ward-identities, which are often written in the 
following form
\begin{equation}\label{cfWard}\begin{aligned}
\big\bra \,T(x)\, & V_{\al_s}(z_s,\bz_s)\dots V_{\al_1}(z_1,\bz_1)\,\big\ket= \\
& \qquad\quad =\;\sum_{i=1}^{s-1}\left(\frac{\De_{\al_i}}{(z-z_i)^2}+
\frac{1}{z-z_i}
\frac{\pa}{\pa z_i}\right)
\big\bra  V_{\al_s}(z_s,\bz_s)\dots V_{\al_1}(z_1,\bz_1)\big\ket.
\end{aligned}
\end{equation}
Validity of the asymptotic relation \rf{classlim} would therefore
imply that 
\begin{equation}\label{C-S}
C_i\;=\;-\frac{\pa}{\pa z_i}S^{\rm cl}[\vf(\underline{x})].
\end{equation}
Equation \rf{C-S} is a nontrivial prediction 
which was proven directly in \cite{CMS,TZ1}. Similar relations can also be
shown to hold in the case of compact Riemann surfaces of arbitrary genus
\cite{ZT}.

\subsection{Geometrical interpretation}

In the case of a generic conformal field theory one usually
interprets the insertion points $z_i$ as parameters for the
``gravitational'' euclidean background on which one studies 
the theory. For the case at hand, however, we may note that
\rf{cleom} implies that the metric 
\begin{equation} \label{metric}
ds^2\;=\;e^{\vf}\,|dz|^2
\end{equation}
represents the unique metric of constant negative curvature that
is compatible with the complex structure on the 
punctured Riemann sphere ${\mathbb P}^1\setminus\{z_1,\dots,z_s\}$.
If one interprets $\vf$ as describing via \rf{metric} the 
gravitational background itself, it becomes natural to study the action 
$ S^{\rm cl}$ as a function of the ``moduli'' $z_1,\dots,z_s$,
thereby elevating the moduli to dynamical variables. 
Indeed, if one fixes only the topological type of Riemann surface 
that one wants to work on (here by choosing the number $s$ 
of operator insertions), then the moduli space  $\CM_s^0$
can be identified with the space of solutions of the 
Liouville equation \rf{cleom}.

In order to formulate the corresponding quantization problem one has
to describe the symplectic structure of the relevant phase space, 
which will here be identified with the space of solutions of 
the Liouville equation \rf{cleom}
on a Riemann surface with fixed topological type. 
Knowing the action $S$ as a function on phase space
makes it possible to extract the corresponding symplectic structure 
in the usual manner. Working with the complex coordinates 
$z_i$, $i=1,\dots,s-3$ it is of course natural to take advantage 
of the complex structure and to define the symplectic form associated
to $S$ as 
\begin{equation}
\omega\;=\;2\pi i\,\pa\bar{\pa}S,
\end{equation}
where $\pa$, $\bar{\pa}$ are the holomorphic and anti-holomorphic
components of the de Rham differential on $\CM_s^0$ respectively.
This symplectic form turns out to be identical to the natural
symplectic form on $\CM_s^0$,
the so-called Weil-Petersson form $\omega_{\rm WP}$ (see e.g. \cite{IT}):
\begin{theorem} \label{TZthm} {\it (Takhtajan-Zograf)\cite{TZ1}}
\begin{equation}
\boxed{\phantom{\sum}\quad\omega\;=\;\omega_{\rm WP}^{}\;\qquad}\;
\end{equation}
\end{theorem}

From this point of view one is naturally led to ask the following
two questions:
\begin{itemize}
\item Is it possible to quantize the spaces
$(\CM_s^0,\omega_{\rm WP})$ in a natural way? In fact, it turns
out to be better to ask for a quantization of the corresponding
{\it Teichm\"uller spaces} $(\CT_s^0,\omega_{\rm WP})$ which are the
universal covering spaces of the moduli spaces 
$\CM_s^0$. The nontrivial topology of the 
moduli spaces $\CM_s^0$ may then be taken into account
by requiring that the covering group 
(the {\it mapping class group} ${\rm MCG}^0_s$)
gets represented by unitary operators.
\item Is it possible to give a natural interpretation for the 
correlation functions in the quantum Liouville theory 
within the quantum theory obtained by quantizing 
$(\CT_s^0,\omega_{\rm WP})$?
\end{itemize}

One might hope
that there exists a ``coherent-state'' representation for the
Hilbert space $\CH_s^0$ in which the wave-functions are
holomorphic functions on the Teichm\"uller spaces
$\CT_s^0$ and the (holomorphic) coordinates $z_n$, 
$n=1,\dots,s-3 $ are realized
as multiplication operators. 
The relations \rf{C-S} furthermore identify
the accessory parameters $C_m$, $m=1,\dots,s-3 $ as some sort of 
conjugate momenta to the holomorphic variables 
$z_n$, in the sense that
\begin{equation}
\{\,z_n,z_m\,\}\,=\,0\,=\,\{\,C_n,C_m\,\},\quad
\{\,z_n,C_m\,\}\,=\,\frac{1}{2\pi i}\de_{nm}\;.
\end{equation}
This suggests that the sought-for 
coherent-state representation should be such that the 
accessory parameters get represented by the holomorphic 
derivatives $\frac{\pa}{\pa z_n}$. The correlation functions
of the quantum Liouville theory, 
being related to the functions $S^{\rm cl}\big[\vf]$
on the phase space $\CT_s^0$
in the semi-classical limit, should correspond to certain 
natural operators $\SO_{\rm A}$ 
on $\CH_s^0$. An operator $\SO$ on $\CH_s^0$
would in a holomorphic representation be represented by a 
kernel $\CK_{\SO}\big(V,\bar{W}\big)$ that depends holomorphically on 
$V=(v_1,\dots,v_{s-3})$ and anti-holomorphically on 
$W=(w_1,\dots w_{s-3})$. 
Could it be that the relation
\begin{equation}\label{question}
\bigg\bra\; V_{\al_s}(\infty,\infty)V_{\al_{s-1}}(1,1)V_{\al_{s-2}}(0,0)
\prod_{i=1}^{s-3} V_{\al_i}(v_i,\bar{w}_i)\;\bigg\ket
\;=\;\CK_{\SO_{\rm A}}\big(V,\bar{W}\big)
\end{equation}
holds for a certain operator $\SO_{\rm A}$? And indeed, the compatibility of 
\rf{question} with the conformal Ward identities   
\rf{cfWard} requires that the operators $\SC_n$ that
correspond to the classical observables $C_n$ should be given by 
\begin{equation}
\SC_n\;=\;b^2\frac{\pa}{\pa z_n}.
\end{equation}

We are going to propose that \rf{question} holds for 
$\SO_{\rm A}={\rm id}$. Unfortunately, so far we only 
know quantization schemes for $\CT_s^0$ in which the wave-functions
are represented as functions of {\it real} variables 
at present \cite{Fo,Ka1,CF,Ka2,Ka3,Ka4}. However, a precise 
relationship between 
these quantization schemes and Liouville theory was 
exhibited in \cite{TT1}, as we will review in the next section
before we further discuss the possible existence of 
a coherent-state representation for the quantized Teichm\"uller spaces
$\CT_s^0$.

\section{Classical and quantized Teichm\"uller spaces}

Throughout this section we will consider Riemann surfaces $\Sigma$
of arbitrary genus $g$ and number of boundary components $s$.

\subsection{The Fenchel-Nielsen coordinates}

A classical set of coordinates for 
the Teichm\"uller spaces $\CT(\Sigma)$ are the so-called 
Fenchel-Nielsen or length-twist coordinates, see
e.g. \cite{IT} for a review. These coordinates
describe the inequivalent ways of gluing hyperbolic 
trinions to form two-dimensional surfaces with negative constant
curvature. 

The basic observation underlying the definition of the 
Fenchel-Nielsen coordinates is the fact that for each triple
$(l_1,l_2,l_3)$ of positive real numbers there is a unique
metric of constant curvature $-1$ on the three-holed sphere (trinion)
such that the boundary components are geodesics with lengths
$l_i$, $i=1,2,3$. A trinion with its metric of constant curvature $-1$
will be called hyperbolic trinion.
There exist three distinguished geodesics 
on each hyperbolic trinion
that connect the boundary components pairwise.

\begin{figure}[htb]
\epsfxsize3cm
\centerline{\epsfbox{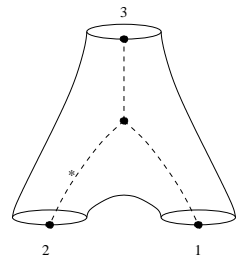}}
\caption{A marked pair of pants}\label{marking}
\end{figure}
Let us call a trinion {\it marked} if it carries 
a graph like the one depicted in Figure \ref{marking}.
Marked trinions can be glued 
such that the markings glue to a three-valent graph on the 
resulting Riemann surface. Conversely one may decompose 
each surface $\Sigma$ of genus $g$ with $s$ 
circular boundaries into trinions by cutting along 
a maximal 
set of mutually non-intersecting cycles $c_i$, $i=1,\dots,\kappa$,
where
\begin{equation}
\kappa\;\equiv\;3g-3+s.
\end{equation}
A trivalent graph $\Ga$ on $\Sigma$  will be 
called Moore-Seiberg graph if it is isotopic to
a graph that can be constructed by gluing marked 
trinions.


Let now $(\Sigma,\Ga)$ be 
a hyperbolic surface with geodesic boundary which is marked with a
Moore-Seiberg graph $\Ga$. 
The Moore-Seiberg graph $\Ga$
defines a decomposition of $\Sigma$ into hyperbolic
trinions by cutting along
mutually non-intersecting geodesics $c_i$, $i=1,\dots,\kappa$. 
The lengths $l_i$, $i=1,\dots,\kappa$ of the geodesics form half of the 
Fenchel-Nielsen coordinates. In oder to define the remaining half let 
us start with the case that the geodesic $c_i$ separates two 
trinions $t_{i,a}$ and $t_{i,b}$. Pick boundary components
$c_{i,a}$ and $c_{i,b}$ of $t_{i,a}$ and $t_{i,b}$ respectively
by starting at $c_i$, following the marking
graphs, and turning left at the vertices. As mentioned above,
there exist distinguished geodesics on $t_{i,a}$ and $t_{i,b}$ 
that connect $c_i$ with $c_{i,a}$ and $c_{i,b}$ respectively. 
Let $\de_i$ be the signed geodesic distance between the end-points
of these geodesics on $c_i$, and let
\begin{equation}
\theta_i\;=\;2\pi\frac{\de_i}{l_i}
\end{equation}
be the corresponding twist-angle. In a similar way one may define 
$\theta_i$ in the case that cutting along $c_i$ opens a handle.

It can be shown (see e.g. \cite{IT}) that the hyperbolic 
surface $\Sigma$ is characterized uniquely by the 
tuple $(l_1,\dots,l_{\kappa};e^{i\theta_1},\dots,e^{i\theta_{\kappa}})\in
(\BR^+)^{\kappa}\ti (S^1)^{\kappa}$. 
In order to describe the Teichm\"uller space $\CT(\Sigma)$ of {\it deformations}
of $\Sigma$ it suffices to allow for 
arbitrary {\it real} values of the twist angles $\theta_i$. 
Points in $\CT(\Sigma)$ are then parametrized by tuples
$(l_1,\dots,l_{\kappa};\theta_1,\dots,\theta_{\kappa})\in
(\BR^+)^{\kappa}\ti \BR^{\kappa}$.

The Weil-Petersson symplectic form becomes particularly simple in terms of
the Fenchel-Nielsen coordinates:
\begin{theorem} {\it (Wolpert)}\cite{Wo}
\begin{equation}
\boxed{\phantom{\sum}\quad\omega_{\rm WP}^{}\;=\;
\sum_{i=1}^{\kappa}\,d\tau_i\wedge dl_i,\quad\tau_i=\frac{1}{2\pi}l_i\theta_i.
\qquad}\;
\end{equation}
\end{theorem}

\subsection{The Moore-Seiberg groupoid}

Changes of the Moore-Seiberg graph
generate canonical transformations from one set of 
Fenchel-Nielsen coordinates to another.
The transitions between different Moore-Seiberg graphs 
on a surface $\Sigma$ generate the
Moore-Seiberg groupoids ${\rm MS}(\Sigma)$.
These groupoids can be characterized
by generators and relations \cite{MS1,BK1}.
In the case that 
$\Sigma$ has genus $g>0$ 
one needs to supplement the generators depicted in Figure \ref{MS1a} 
by one additional generator only, which is shown
in Figure \ref{MS3} below. 
\begin{figure}[h]
\centerline{\epsfxsize7cm\epsfbox{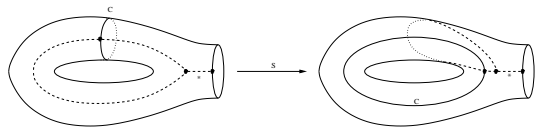}}
\caption{The S-move}\label{MS3} \end{figure}

All relations of the Moore-Seiberg groupoid can be
shown to follow from a finite set of basic relations \cite{MS1,BK1}.
In order to derive the basic relations it suffices to consider 
the cases $g=0$, $s=4,5$ and $g=1$, $s=1,2$. 

\begin{figure}[htb]
\centerline{\epsfxsize7cm\epsfbox{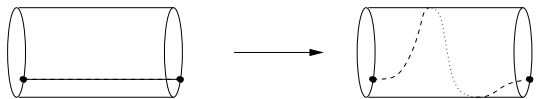}}
\caption{Action of a Dehn-twist on an annulus}\label{dehn} \end{figure}

The Moore-Seiberg groupoid ${\rm MS}(\Sigma)$
contains the {\it mapping class group} ${\rm MCG}(\Sigma)$
as an important sub{\it group}. ${\rm MCG}(\Sigma)$ is generated by the 
Dehn-twists, which represent the operation of cutting out an 
annular region, twisting one end by an angle of $2\pi$ before 
re-gluing, as indicated in Figure \ref{dehn}.
This operation will map any Moore-Seiberg graph
on a surface $\Sigma$ into another one. The action of a Dehn-twist 
on the Moore-Seiberg graphs can be represented by a sequence of the 
elementary transformations depicted in  Figures \ref{MS1a} and
\ref{MS3}.

\subsection{Quantization of the Teichm\"uller spaces}

The quantization of $\CT(\Sigma)$ with the Weil-Petersson 
symplectic form $\omega_{\rm WP}$ is {\it not} canonical
in the Fenchel-Nielsen coordinates. One needs to implement the 
restriction that the operators $\sll_i$, $i=1,\dots,\kappa$, 
which represent the lengths of the closed geodesics $c_i$, have
positive spectrum.
Fortunately there exists an alternative set of coordinates,
introduced by Penner \cite{Pe}, which has the 
advantage to make the quantization of $\CT(\Sigma)$ canonical
\cite{Fo,Ka1,CF}. The result of these constructions is 
an assignment (``modular functor'') 
\begin{equation}\label{modfun}
\Sigma\;\longrightarrow\;\big(\CH^T(\Sigma),\CA^T(\Sigma),
\Pi^T_{\rm\sst MCG}(\Sigma)\big),
\end{equation}
where $\CH^T(\Sigma)$ is a Hilbert space, $\CA^T(\Sigma)$ is an algebra
of operators on $\CH^T(\Sigma)$ that quantizes the commutative 
algebra of functions on $\CT(\Sigma)$, and $\Pi^T_{\rm\sst MCG}(\Sigma)$ 
is a representation
of the mapping class group ${\rm MCG}(\Sigma)$
by unitary operators on $\CH^T(\Sigma)$.

Moreover, there is a construction \cite{Fo} 
that allows one to obtain reasonably simple expressions for
the length functions in terms of the Penner coordinates.
Based on this observation it becomes possible to 
construct \cite{Fo,Ka3,TT1,TT2} the quantum operators 
$\sll_i$, $i=1,\dots,\kappa$ 
that correspond to the geodesic length functions $l_i$.
The key technical result concerning the operators 
$\sll_i$ was obtained in \cite{Ka3,Ka4,TT2}. They indeed 
represent {\it positive} self-adjoint operators with spectrum
$\BR^+$.

It is then not difficult to show that the 
length operators $\sll_c$ and $\sll_{c'}$
associated to two non-intersecting closed
geodesics $c$ and $c'$ always commute, $[\sll_c,\sll_{c'}]=0$.
Specifying a Moore-Seiberg graph $\Ga$ on $\Sigma$ 
amounts to picking a maximal 
set $\{c_1,\dots,c_{\kappa}\}$ 
of mutually non-intersecting closed
geodesics. By simultaneous diagonalization of the 
corresponding length operators $\{\sll_1,\dots,\sll_{\kappa}\}$
one may construct a basis for $\CH^T(\Sigma)$ which consists
of generalized eigenfunctions of $\{\sll_1,\dots,\sll_{\kappa}\}$.

This also allows us to generalize the definition of the 
``modular functor'' \rf{modfun} to the case that 
$\Sigma$ is a Riemann surface with $s$ geodesic boundaries
of fixed length rather than punctures. 
We will use the notation $\Sigma_{L}$ 
to indicate the dependence on the values 
$\Lambda=(\lambda_1,\dots,\lambda_s)$ of the boundary lengths.
It will also be convenient to denote by 
$\Ga_{\Lambda,L}$, with $L=(l_1,\dots,l_{\kappa})\in(\BR^+)^{\kappa}$, 
the Moore-Seiberg graph ``colored'' by assigning the values 
$l_i$ to the geodesics $c_i$.

\begin{theorem} \cite{TT2}
For each Moore-Seiberg graph $\Ga$ on a surface $\Sigma$
of genus $g$ with $s$ geodesic boundaries
there exists a basis 
$\FB_{\Ga}=\{|\Ga_{\Lambda,L}\ket; L\in(\BR^+)^{\kappa}_{}\}$, $\kappa=3g-3+s$,
of generalized eigenfunctions of $\{\sll_1,\dots,\sll_{\kappa}\}$
such that the completeness relation for $\FB_{\Ga}$ can be written as
\begin{equation}\label{idrepr}
\id_{\CA^T(\Sigma)}^{}
\;=\; \int\limits_{(\BR^+)^{\kappa}_{}}
dL\,M(L)\;|\Ga_{\Lambda,L}\ket\bra \Ga_{\Lambda,L}|\,,
\end{equation}
where the measure $M(L)$ is  defined as
\begin{equation}\label{measure}
M(L)\;=\;
\prod_{i=1}^{\kappa}\,\frac{1}{\pi b}\sinh\bigg(\frac{l_i}{2}\bigg)
\sinh\bigg(\frac{l_i}{2b^2}\bigg).
\end{equation}
\end{theorem}

The outcome of these constructions may be considered as 
a quantization of the Fenchel-Nielsen coordinates associated to
a fixed Moore-Seiberg graph $\Ga$. One may
consider the operators $\sll_i$, $i=1,\dots,\kappa$ 
as a natural set of Hamiltonians, and the corresponding one-parameter
groups $e^{-\frac{i}{\hbar}\tau_i\sll_i}$, $\hbar=b^2$ 
as quantum counterparts of the
Fenchel-Nielsen twist flows.

\subsection{Representation of the Moore-Seiberg 
groupoid on $\CH^{\rm T}(\Sigma)$} 

In order to characterize a 
representation of the Moore-Seiberg groupoid it suffices 
to specify the operators 
$\SU_{[\Gamma_2\Gamma_1]}^{}$ 
for the cases where $\Gamma_2$ and $\Gamma_1$
differ by an A-, B- or S-move. The
corresponding operators will be denoted by 
$\SA$, $\SB$ and $\SS$, and will be defined below.\\[1ex]
{\sc A-Move}: In order to describe the representation of the
A-move let $\Sigma_{\rm \Lambda}$ be the 
four-punctured sphere, with parameters ${\rm \Lambda}=(l_4,\dots,l_1)$
associated to the four boundary components 
respectively. The basis
corresponding to the Moore-Seiberg graph $\Ga^s$ depicted in the
leftmost diagram in Figure \ref{MS1a} will be 
denoted by
$\CB_s\equiv\{ |\Ga^s_{\sst {\rm \Lambda},l_s}\ket;l_s\in\BR^+\}$, 
whereas the graph on the second diagram from the
left in Figure \ref{MS1a} will be denoted by $\Ga^t$, 
with corresponding basis
$\CB_t\equiv\{ |\Ga^t_{\sst {\rm \Lambda},l_t}\ket;l_t\in\BR^+\}$. 
The A-move is then represented as an integral
transformation of the following form.
\begin{equation}
|\Ga^s_{\sst {\rm \Lambda},l_s}\ket\;=\;\int\limits_{0}^{\infty}dl_t \,m(l_t)\;
F_{l_sl_t}^{\rm T}\big[\,{}_{l_4}^{l_3}\;{}_{l_1}^{l_2}\,\big]
\;|\Ga^t_{\sst {\rm \Lambda},l_t}\ket\; .
\end{equation}
The measure $m(l)$ is defined by specializing \rf{measure} to $\kappa=1$. 
The kernel 
$F_{l_sl_t}^{\rm T}\big[\,{}_{l_4}^{l_3}\;{}_{l_1}^{l_2}\,\big]$
turns out to be the coincide with
$F_{\al_s\al_t}^{\rm\sst Liou}
\big[\,{}_{\al_4}^{\al_3}\;{}_{\al_1}^{\al_2}\,\big]$
provided that the parameters are related as
\begin{equation}\label{alrel}
\al_\flat\;=\;\frac{Q}{2}+i\frac{l_\flat}{4\pi b},\quad{\rm for}\;\;
\flat\in\{s,t,1,2,3,4\}.
\end{equation}
{\sc B-Move}: The B-move is realized simply by the multiplication 
with the phase factor
\begin{equation}
B^{\rm\sst T}(l_3,l_2,l_1)\;=\;e^{\pi i(\De_{\al_3}-\De_{\al_2}-
\De_{\al_1})},
\end{equation}
where $a_i$ and $l_i$, $i=1,2,3$ are related as in \rf{alrel}, and 
$\De_{\al_i}=\al_i(Q-\al_i)$.\\[1ex]
{\sc S-Move:} Let $\Sigma_l$ be 
a torus with one hole of length $l_e$.
The elements of bases
corresponding to the Moore-Seiberg graphs $\Ga^a$ and $\Ga^b$
shown on left and right halfs of Figure \ref{MS3}
will be denoted by
$\CB_a\equiv\{ |\Ga^a_{l_e,l_a}\ket;l_a\in\BR^+\}$
and
$\CB_b\equiv\{ |\Ga^b_{l_e,l_b}\ket;l_b\in\BR^+\}$
respectively. 
The S-move is then represented as an integral
transformation of the following form.
\begin{equation}
|\Ga^a_{l_e,l_a}\ket\;=\;
\int\limits_{0}^{\infty}dl_b \,m(l_b)\;
S_{l_al_b}(l_e)
\;|\Ga^b_{l_e,l_b}\ket\; .
\end{equation}
The kernel $S_{l_al_b}(l_e)$ is given by the 
following expression:
\begin{equation}
S_{l_al_b}(l_e)=\frac{2^{\frac{3}{2}}}{s_b(p_e)}
\int\limits_{\BR}\!dr\,\prod_{\ep=\pm}
\frac{s_b\big(p_b+\frac{1}{2}(p_e+c_b)+\ep r\big)}
     {s_b\big(p_b-\frac{1}{2}(p_e+c_b)+\ep r\big)}\,e^{4\pi ip_a r}.
\end{equation}
We have set 
$l_{\flat}=4\pi bp_\flat$
for $\flat\in\{a,b,e\}$. 

\begin{theorem}\cite{TT2} \label{mainthm}\begin{itemize}
\item[(i)] The operators $\SA$, $\SB$ and $\SS$ generate  
projective representations $\Pi_{\rm\sst MS}(\Sigma)$ 
of the Moore-Seiberg groupoids, with 
central extension\footnote{In the sense of \cite{BK2}, \S5.7} 
$\ga$ being related to the Liouville central charge
$c=1+6Q^2$ by $\ga=e^{\frac{\pi i}{2}c}$.
\item[(ii)] The corresponding representations 
$\Pi_{\rm\sst MCG}(\Sigma)$ of the mapping class 
groups reproduce the classical 
action of ${\rm MCG}(\Sigma)$ on the 
Teichm\"uller spaces $\CT(\Sigma)$ in the limit $b\ra 0$.
\item[(iii)] The representations $\Pi^0_s$ 
coincide with the representations of the Moore-Seiberg groupoids 
on the spaces of genus zero conformal blocks in Liouville theory.
\end{itemize}
\end{theorem}

\section{Towards a coherent state quantization of $\CT^0_s$}

We now want to further discuss the possible existence of the 
coherent state quantization of $\CT^0_s$. To begin with,
we should formulate more precisely what we consider to be 
defining properties of such a quantization scheme.

The most basic requirement is of course that the quantum states in this
representation are represented by holomorphic functions on $\CT^0_s$, 
and that the quantum observables which correspond to 
{\it analytic} functions on $\CM^0_s$ are represented as
multiplication operators.
We have furthermore argued that the quantum operators $\SC_i$ that
correspond to the accessory parameters $C_i$ 
should be represented as 
\begin{equation}\label{canquant}
\SC_i\;=\;b^2\frac{\pa}{\pa z_i}
\end{equation}
if we use the coordinates $z_i$, $i=1,\dots,s-3$,
as local coordinates
for $\CT^0_s$. 

Another important ingredient must of course be 
the representation of the mapping class groups ${\rm MCG}^0_s$.
Assume that we describe the states $|\psi\ket$ 
by multi-valued wave-functions $\psi(Z)$, 
$Z=(z_1,\dots,z_{s-3})$. For each element $m\in{\rm MCG}^0_s$
and each function $\psi(Z)$
we may define the function $\psi_m(Z)$ via analytic 
continuation as the result of the
covering transformation corresponding to $m$.
In the previous section we have
associated a unitary operator $\SU_{m}$ to each element $m\in{\rm MCG}^0_s$. 
It is then natural to require that
\begin{equation}\label{monoact}
\big(\SU_{m}\psi\big)(Z)\;=\;\psi_m(Z).
\end{equation}
Let us now formulate the
conjecture that we want to propose.

\begin{conjecture}
\begin{itemize}
\item[(i)] There exists a representation 
for the quantized Teichm\"uller
spaces with the properties above. This requires in particular
the existence of a measure $d\si(X)$ on $\CT^0_s$ such that
\[ \bra \psi_2|\psi_1\ket\;=\;\int_{\CT^0_s}d\si(X) \,\big(\Psi_2(X)\big)^*
\Psi_1(X).
\]
We have denoted by $\Psi_i(X)$, $i=1,2$ the analytic functions 
on  $\CT^0_s$ that correspond to the multi-valued 
wave-functions $\psi_i(Z)$.
\item[(ii)] The Liouville conformal blocks $\CF_{S,A}(Z)$ represent the
(generalized) eigenfunctions $\Psi_{\Lambda,L}(Z)=
(Z|\Ga_{\Lambda,L}\ket$ of the length operators in the coherent 
state representation, 
where the sets of parameters are
related via  \rf{alrel}.
\item[(iii)] The vacuum expectation values of primary fields in 
quantum Liouville theory represent the kernel of the identity operator
in the coherent state representation.
\end{itemize}
\end{conjecture}

One may immediately observe that point (iii) of Theorem \ref{mainthm}
strongly supports part (ii) of our conjecture. Indeed, our requirement
\rf{monoact} fixes the monodromies of the 
wave-functions that might represent the eigenfunctions of the 
length operators. We are therefore dealing with a 
Riemann-Hilbert type problem, for which point (iii) of Theorem \ref{mainthm}
asserts the existence of a solution.

Let us furthermore observe that parts (i) and (ii) of our conjecture
actually imply part (iii). This becomes clear if one notes that 
in the length-representation for $\CH^0_s$ one may represent
the identity as in \rf{idrepr}.
Comparing \rf{idrepr}
with the holomorphically factorized representation \rf{s-point2} 
for the vacuum expectation values of primary fields
immediately yields part (iii) of our conjecture.

\subsection{Quantization of the boundaries of the Teichm\"uller spaces}

In the following we will consider surfaces $\Sigma$ for which
all boundary components are punctures, i.e. holes of zero
size.
We want to show that the conjecture above can be verified quite
explicitly if one restricts attention to the behavior of the 
relevant objects near the boundaries of the Teichm\"uller 
spaces which correspond to degenerating Riemann surfaces. 
This will allow us to show that the Liouville conformal blocks 
$\CF_{S,A}(Z)$ are in fact the only reasonable candidates
for the eigenfunctions $\Psi_{\Lambda,L}(Z)$ 
of the length operators as conjectured in part (ii) of our conjecture.

The relevant degenerations correspond to vanishing
of the length $l$ of a closed geodesic $\ga$. Let us
denote the (possibly disconnected) Riemann 
surface obtained by cutting $\Sigma$ along $\ga$ 
by $\Sigma^{\dagger\ga}$.
There always exists
an annular region around the geodesic $\ga$  
that may be modeled by  $A_q=\{zw=q;|z|,|w|<1\}$, where $|q|<1$.
The complex parameter $q$ represents the ``sewing'' modulus
that appears if one reconstructs $\Sigma$ from $\Sigma^{\dagger\ga}$
as in \S\ref{sewing}, with $|q|\ra 0$ corresponding to 
shrinking the length $l\ra 0$. 

The behavior near
degeneration is universal, allowing one to consider $q$ 
independently from the
other moduli of $\Sigma$.
Fortunately it is possible to calculate the asymptotic 
behavior for $|q|\ra 0$ of all relevant objects explicitly.
First, the relation between $|q|$ and the Fenchel-Nielsen
coordinates $(l,\tau)$ associated to $\ga$
is given by \cite{Wo2}
\begin{equation}\label{qFN1}
l=\frac{2\pi^2}{\log(1/|q|)},\qquad 2\pi\frac{\tau}{l}={\rm arg}(q).
\end{equation}
The Weil-Petersson symplectic form $\omega_{\rm WP}$ can be written as 
\cite{Wo2}
\begin{equation}\label{qFN2}
\omega_{\rm WP}\;=\;d\tau\wedge dl\;=\;
i\,\frac{\pi^3}{\log^3(1/|q|)}\,
\frac{dq\wedge d\bar{q}}{2|q|^2}.
\end{equation}
The accessory parameter $C_q$ 
corresponding to $q$ is finally given by the expression \cite{Zo}
\begin{equation}\label{qFN3}
C_q(q,\bar{q})\;=\;
\frac{1}{4q}\biggl(\frac{\pi^2}{\log^2(1/|q|)}-1\biggr).
\end{equation}
By using \rf{qFN1}-\rf{qFN3}
it is straightforward to verify that
\begin{equation}
\{\,q\, ,\, C_q\,\}\;=\;\frac{1}{2\pi i}.
\end{equation}
It is therefore indeed natural to define the quantum operator $\SC_q$
that corresponds to $C_q$ by $b^2 \frac{\pa}{\pa q}$, as required in 
\rf{canquant}.
Let us furthermore note the relation
\begin{equation}
qC_q(q,\bar{q})\;=\;\left(\frac{l}{4\pi}\right)^2-\frac{1}{4},
\end{equation}
which follows from \rf{qFN1} and \rf{qFN3}. We propose to ``quantize'' this
relation as 
\begin{equation}\label{quantprop}
q\frac{\pa}{\pa q} 
\;=\;\left(\frac{\sll_\ga}{4\pi b}\right)^2-\frac{Q^2}{4},
\end{equation}
where $\sll_\ga$ is the geodesic length operator.
The motivation for this particular operator ordering will be 
discussed in Remark 1 below.
It is now of course easy to find the eigenfunctions of 
$\sll_\ga$ in the q-representation. They are given as
\begin{equation}\label{asymefs}
\psi_l(q)\;=\;q^{p^2-\frac{1}{4}Q^2}, \qquad p=\frac{l}{4\pi b}.
\end{equation}
This coincides precisely with the
asymptotic behavior of the Liouville conformal
blocks for $|q|\ra 0$.

\begin{rem}
Instead of \rf{quantprop} one might consider the more general ansatz
\begin{equation}\label{quantprop2}
\nu q\frac{\pa}{\pa q}+(1-\nu)\frac{\pa}{\pa q}q 
\;=\;\left(\frac{\sll_\ga}{4\pi b}\right)^2-\frac{1}{4b^2},
\end{equation}
which parametrizes the ambiguity of ordering the operators $q$ and
$\frac{\pa}{\pa q}$ if we assume that $0\leq \nu\leq 1$.
The choice $\nu=\frac{1}{4}(2-b^2)$ 
adopted in \rf{quantprop} is the 
only one that is compatible with our requirement \rf{monoact},
which determines the monodromy around $q=0$. 
We conclude that the Liouville conformal blocks $\CF_{S,A}(Z)$ 
are in fact the {\it only} candidates
for the  eigenfunctions $\Psi_{\Lambda,L}(Z)$ that are compatible with
both our requirement \rf{monoact} and with the ansatz \rf{quantprop2},
which determines the asymptotic behavior of eigenfunctions of the 
length operator $\sll_{\ga}$ near the boundary of 
the Teichm\"uller spaces.
\end{rem}

Let us finally briefly comment on the existence of a suitable
measure of integration for the definition of the scalar product.
We will propose to consider an ansatz of the form
\begin{equation}
\bra \psi_{2}|\psi_1\ket\;=\;
\int_{\BH} d^2 x\;\nu(x,\bx)\,\big(\psi_{2}(e^{2\pi i x})\big)^*
\psi_{1}(e^{2\pi i x}),
\end{equation}
where we have introduced the ``uniformizing'' variable $x$ via
$q=e^{2\pi ix}$, and integrate $x$ over the upper half plane $\BH$.
In order to satisfy part (ii) of our conjecture we must have
\begin{equation}\label{orthrel}
\int_{\BH} d^2 x\;\nu(x,\bx)\,\big(\psi_{l}(e^{2\pi i x})\big)^*
\psi_{l'}(e^{2\pi i x})\;=\;\frac{1}{m(l)}\de(l-l').
\end{equation}
Taking into account the explicit form \rf{asymefs} of 
$\psi_l(q)$ one may observe that it suffices to assume that
$\nu(x,\bx)$ does not depend on $\Re(x)$ in order to produce the 
delta-distribution in \rf{orthrel}.
$\nu(x,\bx)\equiv\nu'(\Im x)$ can then be determined in terms of 
$m(l)$ by means of an inverse Laplace transformation.

\section{``Exercises''}

\begin{enumerate}

\item Prove Conjecture 5.1.

\item Construct the Liouville conformal blocks on
higher genus Riemann surfaces and prove Conjecture 5.1
for these cases as well.

\item Develop the theory of Teichm\"uller spaces for 
Riemann surfaces with boundaries of arbitrary shape.
Quantize these spaces. Thereby gain insight 
into the relations between the conformal Ward identities,
the geometric action of the Virasoro algebra on moduli
spaces \cite{Ko,BS} and the quantization of Teichm\"uller spaces.

\item Is it possible to quantize the universal
Teichm\"uller space of Bers in a natural way? How is this
related to the solution of Exercise 3?
\end{enumerate}

\end{document}